\documentclass[fleqn,twoside]{article}
\usepackage{espcrc2}
\newcommand{\beq}{\begin{equation}}
\newcommand{\eeq}{\end{equation}}
\newcommand{\bea}{\begin{eqnarray}}
\newcommand{\eea}{\end{eqnarray}}

\newcommand{\gsim}{\lower.7ex\hbox{$
\;\stackrel{\textstyle>}{\sim}\;$}}
\newcommand{\lsim}{\lower.7ex\hbox{$
\;\stackrel{\textstyle<}{\sim}\;$}}

\def\lsim{\mathrel{\rlap{\lower3pt\hbox{\hskip0pt$\sim$}}
    \raise1pt\hbox{$<$}}}         
\def\gsim{\mathrel{\rlap{\lower4pt\hbox{\hskip1pt$\sim$}}
    \raise1pt\hbox{$>$}}}         

\newcommand{\bibit}[1]{\bibitem{#1}}

\newcommand{\aver}[1]{\langle #1\rangle}

\newcommand{\La}{\overline{\Lambda}}
\newcommand{\Lam}{\Lambda_{\rm QCD}}

\newcommand{\GeV}{\,\mbox{GeV}}
\newcommand{\MeV}{\,\mbox{MeV}}

\newcommand{\state}[1]{|#1\rangle}

\newcommand{\vep}{\varepsilon}

\newcommand{\AmS}{{\protect\the\textfont2
  A\kern-.1667em\lower.5ex\hbox{M}\kern-.125emS}}

\title{Strong interaction effects in semileptonic $B$ 
decays\thanks{Talk given at ICHEP-2002, July 24-31 2002, Amsterdam.}}

\author{Nikolai Uraltsev 
{~INFN, Sezione di Milano, Milan, Italy;\\
Department of Physics, University of Notre 
Dame du Lac, Notre Dame, IN 46556, U.S.A.;\\ 
St.\,Petersburg Nuclear Physics Institute, Gatchina, St.\,Petersburg 
188300, Russia}}

\begin{document}
\thispagestyle{empty}

\mathindent=0pt

\begin{abstract}
Strong interaction effects are addressed in connection to extracting
$|V_{cb}|$. A comprehensive approach 
is described not relying on a $1/m_c$ expansion; it allows 
a percent accuracy without ad hoc assumptions about higher-order
effects. An alternative to the $M_X^2$ variable is proposed improving
convergence. Intrinsic hardness of integrated observables 
with a cut on $E_\ell$ is discussed; it can be responsible for the 
behavior of $\aver{M_X^2}$ reported by BaBar.  
Consequences of the proximity to the `BPS' limit are considered.
\vspace*{-6.9cm}\\
\begin{flushright}
{\normalsize
Bicocca-FT-02-20\\
UND-HEP-02-BIG\hspace*{.08em}10\\
hep-ph/0210044\vspace*{4.8cm}}\\
\end{flushright}

\vspace{1pc}
\end{abstract}

\maketitle

The heavy quark (HQ) expansion, a first-principle QCD application of the
Wilsonian OPE to heavy quarks has yielded novel insights into the
dynamics of
heavy flavor hadrons. (For a review, see \cite{ioffe}
and references therein.) 
An important phenomenological application of
the  heavy quark expansion is extracting 
$|V_{\!cb}|$ and $|V_{\!ub}|$ from
measured decay rates with high accuracy and 
little model dependence. This requires 
a genuine control over nonperturbative effects in $B$
decays.

A popular method to determine $|V_{cb}|$ uses the
decay rate $B\!\to\! D^* \,\ell\nu$ near zero recoil. At this kinematic
point the $B\!\to\! D^*$ formfactor $F_{D^*}(0)$ is unity when 
$m_b, m_c \to \infty$. 
Driven by the charm mass scale, power corrections are 
still significant: 
$F_{D^*}\!\simeq\! 0.9$ to order $1/m_Q^2$ \cite{vcb}, and the $1/m_c^3$
effects were estimated to be in the $3\%$ range \cite{beauty95rev}.
Relying on expansion in $1/m_c$ makes it difficult to overcome a $5\%$
level of accuracy here
without compromising reliability of theoretical predictions.

These estimates were supported by recent lattice studies which
yielded, as central values, surprisingly close numbers,
$F_{D^*}\!\simeq\! 0.88$  and $F_{D^*}\!\simeq\! 0.91$ to order $1/m_Q^2$ and
$1/m_Q^3$, respectively \cite{sim}. Since the method is based
on $1/m_Q$ expansion for both $b$ and $c$, an important issue is
higher-order as well as exponential in $m_c$ terms. This
sophisticated lattice approach will hopefully
be refined in the future. Presently a large 
fraction of the corrections to $F_{D^*}(0)\!=\!1$ is still added
theoretically rather than emerges directly in the lattice
simulations.

On experimental side, extrapolating the decay amplitude to zero
recoil introduces additional 
uncertainty. (A parametrization
relied upon is alleged to be rigorously derived from QCD. This is not
true.)
It can be reduced incorporating  the
model-independent inequalities for the slope of the IW function
stemming from the set of the HQ sum rules; however this has not yet been
implemented in the experimental analyses.
\vspace*{1mm}

\noindent 
{\bf Inclusive decays and HQ expansion.}~\hfill
More extensive opportunities are provided by inclusive semileptonic
$B$ decays. Total semileptonic decay rate $\Gamma_{\rm sl}(B)$ is now
one of the best measured quantities in $B$ physics. Theory-wise
nonperturbative effects are controlled by the 10 year old QCD theorem
\cite{qcdtheor} which established absence of the leading $\Lam/m_b$
power corrections to total decay rates. It applies to all
sufficiently inclusive decay probabilities, not only
semileptonic. 
Moreover, the theorem relates the inclusive $B$ widths
to (short-distance) quark masses and expectation values of local
$b$-quark operators in actual $B$ mesons. The general expansion
parameter for inclusive decays is energy release, in $b\to c\,\ell\nu$
it constitutes $m_b\!-\!m_c\simeq 3.5\GeV$.

Heavy quark masses are full-fledged QCD parameters entering various
hadronic processes. The expectation values like $\mu_\pi^2$
determining the width in the OPE likewise enjoy the status of observable
parameters. As pointed out shortly after, the masses and
relevant nonperturbative parameters  can be
determined from the $B$ decay distributions 
themselves \cite{motion,optical}. Nowadays this
strategy is being implemented in a number of experimental studies.

The new generation of data provides accurate measurements of
many inclusive characteristics in $B$ decays. 
It is also
encouraging that proper theoretical formalism gradually finds its way
into their analyses. Recent theoretical findings
allow to shrink theoretical uncertainties -- among them
constraints from the exact HQ sum rules and the consequences of the
proximity to the so-called `BPS' regime signified by the hierarchy
$\mu_\pi^2\!-\!\mu_G^2 \ll \mu_\pi^2$ suggested by experiment.

Present theory allows to aim a percent accuracy in $|V_{cb}|$.
Such a precision becomes possible due to a number of theoretical
advances. 
The low-scale running masses $m_b(\mu)$, $m_c(\mu)$, the
expectation values $\mu_\pi^2(\mu)$, $\mu_G^2(\mu)$... are completely
defined and can be determined from experiment with an in principle
unlimited accuracy. Violation of local duality potentially
limiting theoretical predictability, has been scrutinized and found 
to be negligibly small in total semileptonic
$B$ widths \cite{vadem}. 
Present-day perturbative technology makes computing  $\alpha_s$-corrections
to the Wilson coefficients of nonperturbative operators feasible.
It is also understood how to treat higher-order power
corrections.

High accuracy can be achieved in a comprehensive approach
where many observables are measured in $B$ decays to extract necessary
`theoretical' input parameters. This can be compared with early days
of the heavy quark expansion  when experiment aimed mainly at measuring 
$\Gamma_{\rm sl}(B)$, while the rest had to be supplied by theory. This
limited the accuracy of $|V_{cb}|$ in the mid 1990s by about $5\%$.

With $b\!\to\! c$ widths depending strongly on $m_b\!-\!m_c$, previous
analyses to some extent relied on expansion in $1/m_c$ since employed 
the relation 
\bea
\nonumber
m_b\!-\!m_c \!\!\!\!& =&\!\!\!
\overline{\!\!M\!}_B\!-\overline{\!\!M\!}_D 
+ \frac{\mu_\pi^2}{2}\! 
\left(\!\frac{1}{m_c}\!-\!
\frac{1}{m_b}\!\right) +\\
& & \mbox{\hspace*{-13mm}~} 
\frac{\rho_D^3\!-\!(\rho_{\pi\pi}^3\!+\!\rho_S^3 )}{4} 
\left(\!\frac{1}{m_c^2}\!-\!\frac{1}{m_b^2}\!\right)+
{\cal O}\!\left(\!\frac{1}{m_Q^3}\!\right).
\label{10}
\eea
Reliability of the $1/m_c$ expansion is however questionable. 
Already for the $1/m_Q^2$ terms above one has $\frac{1}{m_c^2}
\!>\!14\frac{1}{m_b^2}$; even for the worst mass scale in the 
width expansion,   
$\frac{1}{(m_b\!-\!m_c)^2}$ is at least $8$ times smaller 
than $\frac{1}{m_c^2}$. 
On top of that there are indications \cite{chrom} that the 
nonlocal correlators
affecting meson masses can be particularly large -- a
pattern also observed in the 't~Hooft model \cite{lebur}. This
expectation is supported by the pilot lattice study \cite{kronsim2}
which -- if taken at face value -- 
suggests a very large value of
$\rho_{\pi\pi}^{3\!}\!\!+\!\rho_{S}^{3}$. 
On the other hand, 
non-local correlators are not measured in inclusive $B$ decays.

A partial cure to this problem 
was suggested recently \cite{chrom}: The proximity to the `BPS' limit 
leads to much smaller power
corrections for the analogue of 
the mass relation (\ref{10}) applied to
ground-state mass difference $M_B\!-\!M_D$ than in the standard 
spin-averaged masses. Since in the conventional
approach this is the major source of uncertainty, {\sf pseudoscalar} 
meson masses 
should be rather used to constrain $m_b\!-\!m_c$.

Many recent extractions of $|V_{cb}|$ from $\Gamma_{\rm sl}(B)$
relied on strong -- and probably unjustified -- assumptions about
``six hadronic $D\!=\!6$ parameters'' appearing in order
$1/m_Q^3$. This led to the lore that the uncertainties in $|V_{cb}|$ are
by far dominated by theory. In fact this depends on the
perspective adopted, and could be traced to the `eclectic' approach where
only $\Gamma_{\rm sl}(B)$, $\La$ and $\mu_\pi^2$ are relegated to
experiment, whereas all the remaining information must come from
theory. Since theory itself warns that the mass relation (\ref{10})
for charm is the weakest point, and a number of rigorous theoretical
constraints are disregarded here, such an approach clearly cries for
improvement.

Fortunately, there is a way totally free from relying on charm mass
expansion; the validity of the latter can rather be examined a
posteriori.
It was put forward some time ago \cite{cernckm} to utilize
the power of the comprehensive approach and makes full use
of a few key facts \cite{optical,motion}:\\
~\hspace*{.2em}$\bullet$ Total width to order $1/m_b^3$ is affected by
a single new Darwin operator (its expectation value is $\rho_D^3$); 
the moments also weakly depend on $\rho_{LS}^3$.\\
~\hspace*{.2em}$\bullet$ No nonlocal correlators ever enter {\it per se}.\\
~\hspace*{.2em}$\bullet$ Deviations from 
the HQ limit in the expectation values are
driven by the full scale $1/m_b$ (and are additionally
suppressed by proximity to the BPS limit); they are negligible in
practice. \\
~\hspace*{.2em}$\bullet$ Exact sum rules and 
inequalities which hold for properly
defined Wilsonian parameters.

Some of the HQ parameters like $\mu_G^2$ are known beforehand. Proper 
field-theoretic definition allows its accurate determination 
from the $B^*\!-\!B$ mass splitting:
$\mu_G^2(1\GeV)\!=\!0.35^{+.03}_{-.02}\GeV^2$ \cite{chrom}. 
A priori less certain is $\mu_\pi^2$. 
However, the inequality $\mu_\pi^2\!>\!\mu_G^2$ valid 
for any definition of
kinetic and chromomagnetic operators respecting the QCD commutation
relation $[D_j,D_k]\!=\!-ig_sG_{jk}$, 
and the corresponding sum rules 
essentially limit its range: $\mu_\pi^2(1\GeV)\!=\!0.45\!\pm\!
0.1\GeV^2$.

Running $b$ quark mass was accurately extracted from 
$\sigma\left(e^+e^-\!\!\to\! \Upsilon(nS)\right)$ in the end of 
the 1990s: 
$m_b(1\GeV)\!=\!4.57\!\pm\! 0.06\GeV$ for the ``kinetic'' $m_b(\mu)$. 
However, considering all available constraints, I think
that $4.57\GeV$ is on the lower side of the $m_b$ range 
centered rather around $4.63\GeV$.

Often extracted from the data are the ``HQET parameters''
($-\lambda_1, \La$) -- they actually correspond to extrapolating the
$\mu$-dependent quantities down to $\mu\!=\!0$. They are ill-defined and
make no sense out of the context of a concrete computation; they are
meaningful only as intermediate stage entries. However,
a translation can often be made into
properly defined parameters. Say,
in the context of the recent CLEO analyses it reads 
\bea
\nonumber
\La_{\rm HQET}\!\!\!&\simeq&\!\! \La(1\GeV)-0.255\GeV \\ 
-\lambda_1\! &\simeq&\!\!\! \mu_\pi^2(1\GeV)-0.18\GeV^2\;.
\label{20}
\eea
The central values quoted by CLEO 
\cite{cleoamst} thus correspond to $m_b(1\GeV)\!=\!4.62\GeV$,
$\;\,\mu_\pi^2(1\GeV)\!=\!0.43\GeV^2$, surprisingly close to the theoretical
expectations!
\vspace*{1mm}

\noindent
{\bf Lepton and hadron moments.}
Moments of the charged lepton energy in the semileptonic $B$ decays
are traditional observables to measure heavy quark parameters. New at this
conference are DELPHI results for the first three moments. 
Two moments with the lower cut at $E_\ell=1.5\GeV$ or $E_\ell=1.7\GeV$
are presented by CLEO, who also measured average photon energy
$\aver{E_\gamma}$ subject to constraint $E_\gamma>2\GeV$. 

Another useful set of observables are moments of the invariant
hadronic mass squared $M_X^2$ in semileptonic decays. Their utility 
follows from the fact \cite{optical} that, at least if
charm were heavy enough  
the first, second and third moments would more or less
directly yield $\La$, $\mu_\pi^2$ and 
$\rho_D^3$.
The hadronic moments were measured in different settings by
DELPHI, CLEO and BaBar. The details can be found in the original
experimental talks \cite{cleoamst,delphiamst,babar,battag}.

Let me now illustrate how this strategy works number-wise.
Lepton energy moments, for instance, are given 
by the following approximate expressions ($b\!\to\! u$ decays are neglected):
\bea
\nonumber
&&\mbox{\hspace*{-8mm}~}\aver{E_\ell}\!=\!1.38\GeV + 0.38[(m_b\!-\!4.6\GeV) 
-\\
\nonumber
&&\mbox{\hspace*{.55mm}~}0.7(m_c\!-\!1.15\GeV)]
+0.03(\mu_\pi^2\!-\!0.4\GeV^2) \\
\nonumber
&&\mbox{\hspace*{28mm}~}-0.09(\tilde\rho_D^3\!-\!0.12\GeV^3)\;,
\rule[-2mm]{0mm}{4mm}\\
\nonumber
&&\mbox{\hspace*{-8mm}~}\aver{(E_\ell\!-\!\aver{E_\ell})^2}\!=\!0.18\GeV^2 \!+ 
0.1[(m_b\!-\!4.6\GeV)\!-\\
\nonumber 
&&\mbox{\hspace*{1mm}~} 0.6(m_c\!-\!1.15\GeV)] +
0.045(\mu_\pi^2\!-\!0.4\GeV^2)\\  
\nonumber 
&&\mbox{\hspace*{28mm}~}-0.06(\tilde\rho_D^3\!-\!0.12\GeV^3)\;,\\
\nonumber 
&&\mbox{\hspace*{-8mm}~}\aver{(E_\ell\!-\!\aver{E_\ell})^3} = -0.033\GeV^3 - 
0.03\,[(m_b\!-\!\\
\nonumber 
&&\mbox{\hspace*{-3mm}~}4.6\GeV)
-0.8(m_c\!-\!1.15\GeV)]  
+ 0.024(\mu_\pi^2\!-\!
\\   
&&\mbox{\hspace*{5.5mm}~}0.4\GeV^2)
-0.035(\tilde\rho_D^3\!-\!0.12\GeV^3)\;.
\label{30}
\eea
The moments depend basically on one and the same
combination of masses $m_b\!-\!0.65m_c$; dependence on $\mu_\pi^2$ is
rather weak. To even larger extent this applies to the CLEO's cut
moments $R_1$, $R_2$ and the ratio $R_0$ -- they depend practically on
a single combination $m_b\!-\!0.63m_c\!+\!0.3\mu_\pi^2$. The effect of
the spin-orbital average $\rho_{LS}^3$ is negligible. 

Now take a look at $|V_{cb}|$. Its value extracted from $\Gamma_{\rm sl}(B)$ 
has the following dependence on
the HQ parameters:{\small \vspace*{-1mm}
\bea
\nonumber
&&\mbox{\hspace*{-8mm}~}\mbox{{\large
$\frac{|V_{cb}|}{0.042}$}} 
\!=\! 1\!-\!0.65[(m_b\!-\!4.6\GeV)
\!-\!0.61(m_c\!-\! 1.15\GeV)] \\
\nonumber
&&\mbox{\hspace*{-5.6mm}~}+0.013\,(\mu_\pi^2\!-\!0.4\GeV^2)
+\,0.1(\tilde\rho_D^3\!-\!
0.12\GeV^3)\\
\nonumber
&&\mbox{\hspace*{-5.6mm}~} +0.06(\mu_G^2\!-\!0.35\GeV^2)
-0.01(\rho_{LS}^3\!+\!0.15\GeV^3) = \rule[-2mm]{0mm}{4mm}\\ 
\nonumber
&&\mbox{\hspace*{-8mm}~}\raisebox{-.2mm}{\mbox{{\normalsize$1$}}}\,-\,
\raisebox{-.2mm}{\mbox{{\normalsize$\frac{0.65}{0.38}$}}}\,
[\aver{E_\ell}\!-\!1.38\GeV] \,-\, 0.06\, (m_c\!-\!1.15\GeV) 
\\
\nonumber
&&\mbox{\hspace*{-2.5mm}~}- 0.07(\mu_\pi^2\!-\!0.4\GeV^2)
-0.05(\tilde\rho_D^3\!-\!0.12\GeV^3)-
\\
&&\mbox{\hspace*{-6mm}~} 0.08(\mu_G^2\!-\!0.35\GeV^2)-
0.005\,(\rho_{LS}^3\!+\!0.15\GeV^3);
\label{34}
\eea}
\hspace*{-.4em}a combination of the parameters has been replaced by 
the first lepton moment in Eq.~(\ref{30}), and the 
sensitivity to $\mu_G^2$ and $\rho_{LS}^3$ is illustrated. 
If we do a similar
exercise for the lepton moment $R_1$ with the cut at
$E_\ell\!>\!1.5\GeV$, the
coefficients giving the remaining sensitivity of $|V_{cb}|$ to 
$m_c$, $\mu_\pi^2$
and $\tilde\rho_D^3$ will become $0.02$, $0.19$ and $0.13$,
respectively.
We see that the precise value of charm mass is
irrelevant, but reasonable accuracy in $\mu_\pi^2$
and $\tilde\rho_D^3$ is required.

The first {\tt hadronic} moment takes the form 
\bea
\nonumber
&&\mbox{\hspace*{-8mm}~}
\aver{M_X^2} = 4.54\GeV^2 - 5.0\,[(m_b\!-\!4.6\GeV)-\\
\nonumber
&&\mbox{\hspace*{32mm}~}
0.62\,(m_c\!-\!1.15\GeV)]\\
&&\mbox{\hspace*{-4.3mm}~}
-0.66\,(\mu_\pi^2\!-\!0.4\GeV^2)+
(\tilde\rho_D^3\!-\!0.12\GeV^3),
\label{40}
\eea
i.e., given by nearly the same combination
$m_b\!-\!0.7m_c\!+\!0.1\mu_\pi^2\!-\!0.2\rho_D^3$ as the lepton moment. 
Not very constraining, it provides, however a highly nontrivial check
of the HQ expansion. These two first 
moments together, for example verify the heavy
quark sum rule for $M_B\!-\!m_b$ with the accuracy about $40\MeV$! 
In this respect it is more elaborate than the higher lepton moments.

The dependence expectedly changes for higher hadronic moments:
\bea
\nonumber
&&\mbox{\hspace*{-8mm}~}
\aver{(M_{X\!}^2\!-\!\aver{M_X^2})^2}\!=\!1.2\GeV^{4\!}  
\!-\!0.003(m_b\!-\!4.6\GeV)\\
\nonumber
&& \mbox{\hspace*{-2mm}~}-0.68\,(m_c\!-\!1.15\GeV) +
4.5\,(\mu_\pi^2\!-\!0.4\GeV^2)\\
\nonumber
&&\mbox{\hspace*{28mm}~} 
-5.5\,(\tilde\rho_D^3\!-\!0.12\GeV^3)\;,\\
\nonumber
&&\mbox{\hspace*{-8mm}~} 
\aver{(M_X^2\!-\!\aver{M_X^2})^3}= 4\GeV^6 +(m_b\!-\!4.6\GeV)
\\
\nonumber
&&\mbox{\hspace*{3mm}~}
-3\,(m_c\!-\!1.15\GeV) + 5\,(\mu_\pi^2\!-\!0.4\GeV^2)
\\
&&\mbox{\hspace*{23mm}~}
+13\,(\tilde\rho_D^3\!-\!0.12\GeV^3)\;.
\label{44}
\eea
Ideally, they would measure the kinetic and Darwin expectation values
separately. At the moment, however, we have only an approximate
evaluation and informative upper bound on $\tilde\rho_D^3$. The
current sensitivity to $\mu_\pi^2$ and $\tilde\rho_D^3$ is about
$0.1\GeV^2$ and $0.1\GeV^3$, respectively.

We see that measuring the second and third hadronic moments is the
real step in implementing the comprehensive program of extracting
$|V_{cb}|$. Clearly, more work -- both theoretical and experimental --
is required to fully use its power.  It is crucial that
this extraction carries no hidden assumptions, and at no point we
rely on $1/m_c$ expansion. Charm quark could be either heavy, or
light as strange or up quark, without deteriorating -- and even
improving the accuracy!
\vspace*{1mm}

\noindent
{\bf Experimental cuts and hardness.}
There is a problem, however, which should not be underestimated. 
The intrinsic `{\tt hardness}' of the moments deteriorates when the cut
on $E_\ell$ is imposed. As a result, say the extraordinary
experimental 
accuracy of CLEO's $R_0$--$R_2$ cannot be even nearly
utilized by theory, whether or not the expressions we use make
this explicit.\footnote{An instructive example of how naive analysis
can miss such effects was given in \cite{uses}, Sect.~5.}

For total widths the effective energy scale parameter is generally
${\cal Q}\!=\!m_b\!-\!m_c$. 
When OPE applies we can go beyond purely qualitative speculations about
hardness. Then 
it is typically given by ${\cal Q}\!\lsim\! \omega_{\rm
max}$, with $\omega_{\rm max}$ the threshold energy at which the
decay process disappears once $m_b$ is replaced by
$m_b\!-\!\omega$.  With the $E_\ell\!>\!E_{\rm min}$ cut then 
\beq
{\cal Q}\simeq m_b-E_{\rm min}-\sqrt{E_{\rm min}^2+m_c^2} 
\label{50}
\eeq
constituting only meager $1.25\GeV$ for 
$E_{\rm min}\!=\!1.5\GeV$, and falls
even below $1\GeV$ for the decays with $E_\ell \!>\! 1.7 \GeV$. 
This may explain the unexpected behavior of the first
hadronic moment with respect to the cut on $E_\ell$ reported by BaBar
earlier at this session \cite{babar}. 

In  $b\!\to\! s\!+\!\gamma$ decays one has ${\cal Q}\!\simeq\!
m_b\!-\!2E_{\rm min}$, once again a rather soft scale $1.2\GeV$ if the
lower cut is set at $E_\gamma\!=\!2\GeV$. Hence, the reliability of theory
can be questioned when one aims for maximum precision.
For higher moments the hardness further deteriorates in
either decays. A high premium then 
should be placed for lowering the cuts \cite{uses}.

On the theoretical side, the higher hadronic moments can be affected
by nonperturbative physics formally scaling as powers of $1/m_b$ 
greater than $3$. At the same time, these moments are instrumental for
a truly model-independent comprehensive studies of $B$
mesons; improvement is needed already for the third moment, its
expression given above is not too accurate. 
Considering alternative
kinematic variables will help to improve the convergence. 
Namely, it is advantageous to trade the traditional
hadronic mass $M_X^2$ for the observable more closely corresponding to
the quark virtuality $\Delta$, defined as 
\beq
{\cal N}_X^2 \!=\!M_X^2\!-\!2\tilde\Lambda E_X\,, 
\label{52}
\eeq
where $E_X\!=\!M_B\!-\!q_0$ is the total hadronic energy in the $B$ restframe, 
and $\tilde\Lambda$ a fixed mass parameter. Its
preferred values are about $M_B\!-\!m_b(1\GeV)$ and can be taken
$600-700\MeV$.
The higher moments 
$\aver{({\cal N}_X^2\!-\!\aver{{\cal N}_X^2})^2}$, $\aver{({\cal
N}_X^2\!-\!\aver{{\cal N}_X^2})^3}$... should enjoy better theoretical
stability.

The kinematic variable ${\cal N}_X^2$ is not well constrained
inclusively at LEP experiments, however can be used in the $B$
threshold production at CLEO and $B$ factories.\footnote{I am
grateful to experimental colleagues for discussing this point.} 
This possibility should be carefully explored.
\vspace*{1mm}

\noindent
{\bf BPS limit.}
An intriguing theoretical environment opens up if
$\mu_\pi^2(1\GeV)$ 
is eventually confirmed to be close enough to
$\mu_G^2(1\GeV)$ as currently suggested by experiment, say it does
not exceed $0.45\GeV^2$. If $\mu_\pi^2\!-\!\mu_G^2
\!\ll \!\mu_\pi^2$ it is advantageous to analyze strong dynamics 
expanding around the point
$\mu_\pi^2\!=\!\mu_G^2$ \cite{chrom}. This is not just 
one point of a continuum in
the parameter space, but a quite special `BPS' limit where the
heavy flavor  
ground state satisfies functional relations 
$\vec{\sigma}\vec{\pi}\state{B}\!=\!0$. This limit is 
remarkable in
many respects, for example, saturates the bound \cite{newsr}
$\varrho^2\!-\!\frac{3}{4}$ for the slope of the IW function. 
In some instances like the $B\!\to\! D$ zero-recoil 
amplitude it extends the heavy flavor (but not spin) symmetry to  
higher orders in $1/m_Q$. One of its  
practical application has been mentioned -- the robust
relation for $m_b\!-\!m_c$ via $M_B\!-\!M_D$. Exclusive $B\!\to\! D^*$
decay can also benefit from the proximity to BPS. The exact spin
sum rules  
yield a constraint on the IW slope
\beq
\mu_\pi^2\!-\!\mu_G^2 \!=\!3\tilde\vep^2
(\varrho^2\!-\!\mbox{$\frac{3}{4}$}),\;\;\;\; 0.45\GeV \!\lsim\!
\tilde\vep \!\lsim\! 1\GeV
\label{56}
\eeq
thus leaving only a small room for the slope of the actual $B\!\to\! D^*$
formfactor, excluding values of $\hat\varrho^2$ 
exceeding $1.15\!-\!1.2$. This would be a
very constraining result for a number of experimental studies.
\vspace*{1mm}

\noindent
{\bf Conclusions.} Experiment has entered a new era of exploring $B$
physics at the nonperturbative level, with qualitative improvement in
$|V_{cb}|$. 
The comprehensive approach will allow to reach a percent level of
reliable 
accuracy in translating  $\Gamma_{\rm sl}(B)$ to $|V_{cb}|$. 
Recent experiments have
set solid grounds for dedicated future studies at $B$
factories. We already observe a nontrivial
consistency between quite different measurements, 
and between experiment and QCD-based theory.

There are obvious lessons to infer. Experiment must strive to weaken
the cuts in inclusive measurements used in extracting $|V_{cb}|$. 
Close attention should be paid to higher moments or their special
combinations, as well as exploring complementary kinematic observables.

The theory of heavy quark decays is now a
mature branch of QCD -- still  there are a number of directions 
where it can be developed further. 
I believe that 
the analyses presented at the Conference
should provide theorists with 
substantial motivation to refine it at
least in  the following:\\
\hspace*{.3em}$\bullet$ Calculating perturbative corrections to Wilson
coefficients of subleading operators.\\
\hspace*{.3em}$\bullet$ Scrutiny of higher-order power
corrections.\\
\hspace*{.3em}$\bullet$ A thorough study of alternative kinematic
variables, for instance moments of ${\cal N}_X^2$.

To fully realize the physical information in the quest for the
ultimate precision, a truly comprehensive analysis must implement all
theoretical constraints on HQ parameters; the suitable framework uses
well-defined running parameters having physical meaning. Heavy quark
sum rules appear to yield strong constraints on the parameter
space; it is important to study the question of their saturation. If
a low $\mu_\pi^2$ around $0.45\GeV^2$ is confirmed by experiment, the
BPS expansion will play an important role in analyzing nonperturbative
effects -- in particular, guide us through higher-order corrections.
\vspace*{1mm}

{\bf Acknowledgments:} It is a pleasure to thank\\ M.\,Artuso, 
M.\,Battaglia,~I.\,Bigi,~M.\,Calvi,~P.\,Gam\-bino, V.\,Luth 
and P.\,Roudeau for 
helpful discus\-sions. 
This work was supported in part by the NSF under grant number PHY-0087419.

\end{document}